\documentclass[twocolumn,showpacs,amsmath,amssymb,nofootinbib,notitlepage,superscriptaddress,jcp,longbibliography]{revtex4-1}

\usepackage[pdftex]{graphicx} % Include figure files
\usepackage{dcolumn}  % Align table columns on decimal point
\usepackage{bm}       % bold math
\usepackage[usenames,dvipsnames]{color}
\usepackage{epstopdf}
\definecolor{URLCOL}{rgb}{0,0.17,0.43} %external link color
\definecolor{LINKCOL}{rgb}{0.05,0.4,0} %internal link color
\definecolor{CITECOL}{rgb}{0.35,0,0.48} %link to bibliography
\usepackage[pdftex,bookmarks,breaklinks,bookmarksopen,bookmarksnumbered,colorlinks,
			linkcolor=LINKCOL,linktocpage,citecolor=CITECOL,urlcolor=URLCOL,
			pdfpagemode=UseOutline,pdftex]{hyperref}

\usepackage{ctable}
\usepackage{multirow}
\usepackage{isomath}

\usepackage{booktabs}
\usepackage{comment}
\usepackage{natbib}
\def\sss{\scriptscriptstyle\rm}

\def\loc{^{\rm loc}}

\def\bn{\vectorsym{n}}

\def\Tv{\vectorsym{T}}
\def\ba{\vectorsym{\alpha}}
\def\TML{T^{\rm ML}}
\def\K{\matrixsym{K}}

\def\W{{\rm ^W}}

\def\K{\matrixsym{K}}

\def\TML{T^{\rm ML}}
\def\ML{^{\rm ML}}

\definecolor{TITLECOL}{rgb}{0.1,0.2,0.7} %title color
\definecolor{SECOL}{rgb}{0.1,0.2,0.7} %sec color
\definecolor{CONTENTSCOL}{rgb}{0.1,0.2,0.7} %can choose the table of contents title to have same color as sec
\definecolor{SSECOL}{rgb}{0.25,0,0.48} %ssection color
\definecolor{SSSECOL}{rgb}{0.2,0.08,0.53} %subsubsection color  0.2,0.08,0.53

\def\sss{\scriptscriptstyle\rm}

\def\bea{\begin{eqnarray}}
\def\eea{\end{eqnarray}}
\def\ben{\begin{equation}}
\def\een{\end{equation}}
\def\benu{\begin{enumerate}}
\def\enu{\end{enumerate}}

\def\bei{\begin{itemize}}
\def\eei{\end{itemize}}
\def\beit{\begin{itemize}}
\def\eit{\end{itemize}}
\def\benu{\begin{enumerate}}
\def\enu{\end{enumerate}}

\def\s{_{\sss S}}

\def\ee{_{\rm ee}}

\def\LDA{^{\rm LDA}}

\begin{document}

\title{Orbital-free Bond Breaking via Machine Learning}

\author{John C. Snyder}
\affiliation{Departments of Chemistry and of Physics,
University of California, Irvine, CA 92697, USA}

\author{Matthias Rupp}
\affiliation{Institute of Pharmaceutical Sciences, ETH Zurich, 8093 Z{\"u}rich, Switzerland}

\author{Katja Hansen}
\affiliation{Fritz-Haber-Institut der Max-Planck-Gesellschaft, 14195 Berlin, Germany}

\author{Leo Blooston}
\affiliation{Department of Chemistry,
University of California, Irvine, CA 92697, USA}

\author{Klaus-Robert M{\"u}ller}
\affiliation{Machine Learning Group, Technical University of Berlin, 10587 Berlin, Germany}
\affiliation{Department of Brain and Cognitive Engineering, Korea University,
Anam-dong, Seongbuk-gu, Seoul 136-713, Korea}

\author{Kieron Burke}
\affiliation{Departments of Chemistry and of Physics,
University of California, Irvine, CA 92697, USA}

\date{\today}

\begin{abstract}
Machine learning is used to approximate the kinetic energy of 
one dimensional diatomics as a functional of the electron density.
The functional can accurately dissociate a diatomic, and can be 
systematically improved with training.
Highly accurate self-consistent densities and molecular forces are found,
indicating the possibility for ab-initio molecular dynamics simulations.

\end{abstract}

\maketitle

%--------------------------------------------------------------------------%

Kohn-Sham density functional theory (KS-DFT) \cite{HK64,KS65} is a widely used electronic
structure method, striking a balance between accuracy and computational
efficiency \cite{B12}.
KS-DFT is not a pure DFT, as it requires solving a self-consistent
set of orbital equations \cite{BW12}.   In return, only a small
fraction of the total energy, the
exchange-correlation (XC) energy, need be approximated as a
functional of the electronic spin densities. This produces far greater
accuracy relative to a pure DFT such as Thomas-Fermi
theory \cite{DG90}.

The computational bottleneck in KS-DFT calculations is the
need to solve the KS equations, which formally scale as $N^3$, where $N$
is the number of electrons.  Thus there is strong interest in 
constructing an orbital-free DFT, avoiding this step \cite{KT12}.
A sufficiently accurate approximation to $T\s[n]$, the kinetic
energy (KE) of KS electrons, would produce an orbital-free scheme,
greatly reducing the computational cost of DFT without sacrificing accuracy.

Many research efforts have recently focused in this direction \cite{kjth2009}.
Unfortunately,
the relative accuracy requirements of a KE functional
are much stricter than those of an XC functional, because the KE
 is typically comparable to the total energy of the system \cite{DG90}.
Worse, we also need accurate functional derivatives, since 
ultimately the density must be determined self-consistently via an Euler equation.
The standard approximations using local and semi-local forms do not
yield accurate derivatives \cite{UG94}.

Approximating $T\s[n]$ has proven to be a difficult task for both extended and finite
systems \cite{KT12}. Some build on Thomas-Fermi theory
with gradient expansions and various mixing coefficients \cite{kjth2009} or on generalized gradient approximations (GGAs) with enhancement factors based on ``conjointness'' \cite{KJTH09b}. Others approximate
$T_\theta = T\s - T\W \geq 0$, where $T\W$ is the von Weizs{\"a}cker KE \cite{vw1935}, or
attempt to produce the correct linear response \cite{KT12}.
For multiple bonds, no present KE functional can accurately describe molecules far from
equilibrium structures nor properly dissociate a diatomic. Moreover, solving the orbital-free Euler equation
for GGA-like functionals can be difficult due to poor functional derivatives near
nuclei \cite{KT12}.

A particularly difficult problem is to
correctly dissociate a chemical bond.  Any locally-based approximation
has difficulties when the bond length is stretched to large distances.
The fragments often contain fractional electron numbers, for which local
approximations yield very wrong answers \cite{CMY08}.
The worst case is
the KE of a single bond in one dimension, where a local approximation yields
fragments energies that are incorrect by a factor of 4 in the stretched limit.

\begin{figure}[b]
\includegraphics[width=8.5cm]{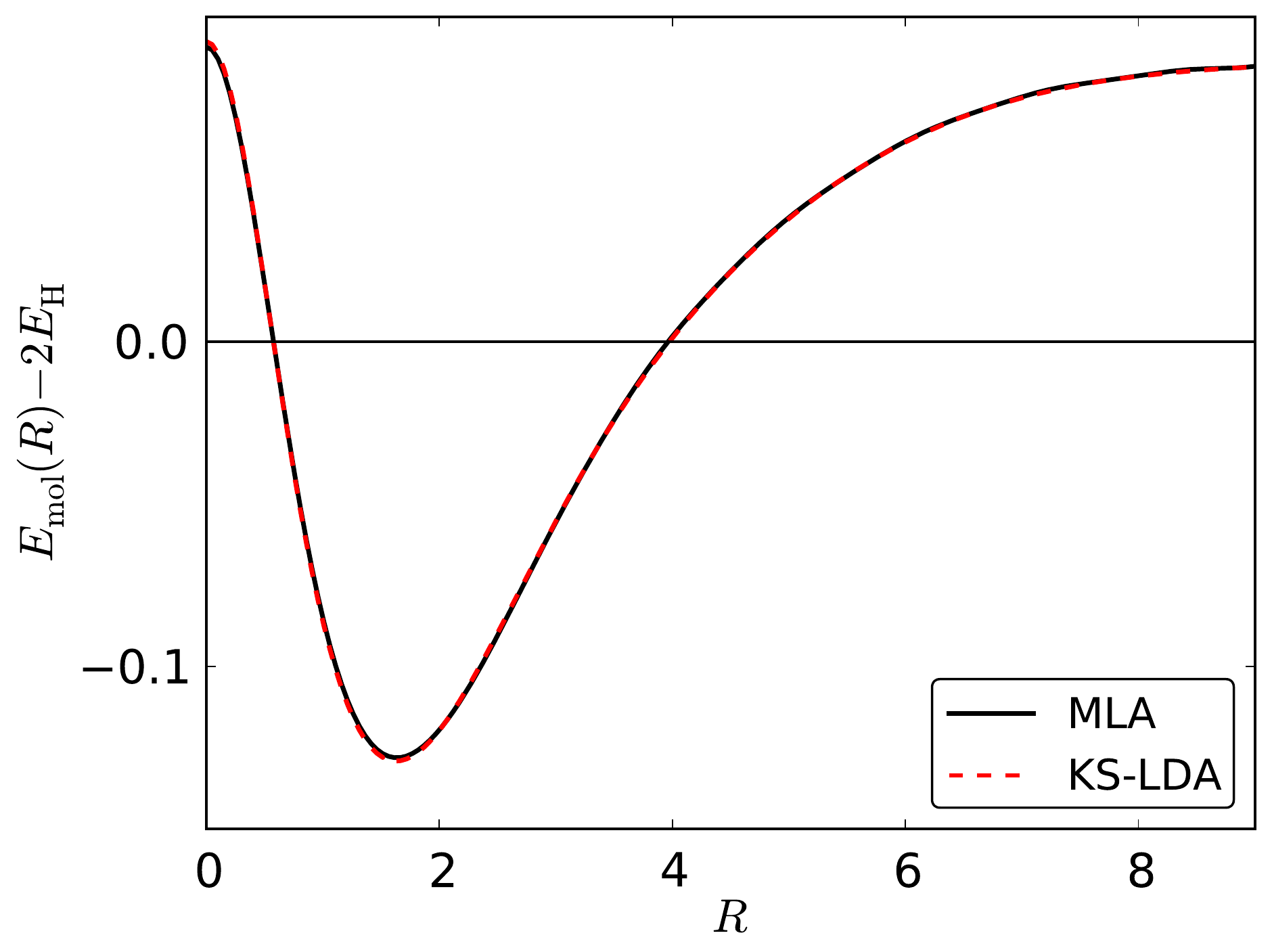}
\caption{The molecular binding curve for a 1d single bond (in atomic units). The machine
learning approximation (MLA) curve is found found self-consistently, using 10
KS-LDA densities and kinetic energies (spaced evenly from $R=0$ to 10) for training.
Note that the incorrect large $R$ limit is due to deficiencies in the LDA XC, {\em not} 
our MLA for the KE.}
\label{f:bond}
\end{figure}

To tackle these difficulties, we turn to machine learning (ML), a
powerful tool for learning high-dimensional patterns via induction that
has been very successful in many applications \cite{mmrts2001}
including quantum chemistry \cite{i2007, BPKC10, RTML12, PHSR12}.
Some of us recently suggested a new paradigm in density functional approximation using
ML to approximate density functionals \cite{SRHM12}. 
In that work, all particles were confined to a box so no densities looked like
those of well-separated atoms.
In Fig.~\ref{f:bond}, we show what happens when we apply ML algorithms self-consistently to
fermions in two wells as they are separated:
ML has no particular difficulty in treating a
situation that is virtually impossible with traditional DFT approximations.
In this work, we construct an orbital-free KE functional based
on ML that is capable of accurately describing 1d diatomics from the united atom limit
to complete nuclear separation. Moreover, we obtain accurate self-consistent densities
 and molecular forces.

We use standard methods from ML \cite{htf2009}, and atomic units throughout.
Using kernel ridge regression (KRR), which is non-linear regression with regularization 
to prevent overfitting \cite{htf2009},
our machine learning approximation (MLA) for the KE is
\ben
\TML(\bn) = \sum_{j=1}^M \alpha_j k(\bn_j, \bn),
\label{eq:TMLdef}
\een
where $\alpha_j$ are weights to be determined, $\bn_j$ are training densities, $M$ is the number
of training densities, and $k$ is the kernel,
which measures similarity between densities.
We choose the Gaussian kernel
\ben
k(\bn, \bn') = \exp(-\| \bn - \bn' \|^2/(2\sigma^2)),
\een
where $\sigma$ is called the length scale.
The weights are found by minimizing the cost function
\ben
{\mathcal C}(\ba) =  \sum_{j=1}^M \Delta T_j ^2 +  \lambda \ba^T \K \ba,
\een
where $\Delta T_j = \TML_j - T_j$, $\ba = (\alpha_1, \dots, \alpha_M)$ and
$\K$ is the kernel matrix, $\K_{ij} = k(\bn_i, \bn_j)$.
The second term is a regularizer that
penalizes large weights to prevent overfitting.
Minimizing ${\mathcal C}(\ba)$ gives
\ben
\ba = (\K + \lambda \matrixsym{I})^{-1} \Tv,
\label{eq:KRRsolution}
\een
where $\matrixsym{I}$ is the identity matrix and $\Tv = (T_1, \dots, T_M)$.
The global parameters $\sigma$ and $\lambda$ are determined through leave-one-out cross-validation:
For each density $\bn'$ in the training set, the 
functional is trained on all densities except for $\bn'$
and $\sigma$ and $\lambda$ are optimized by minimizing the absolute error on
the $\bn'$. Final values are chosen as the median over all optimum values.
To test performance, the functional is always evaluated on new densities not
in the training set.

 \begin{figure}[t]
\includegraphics[width=8.5cm]{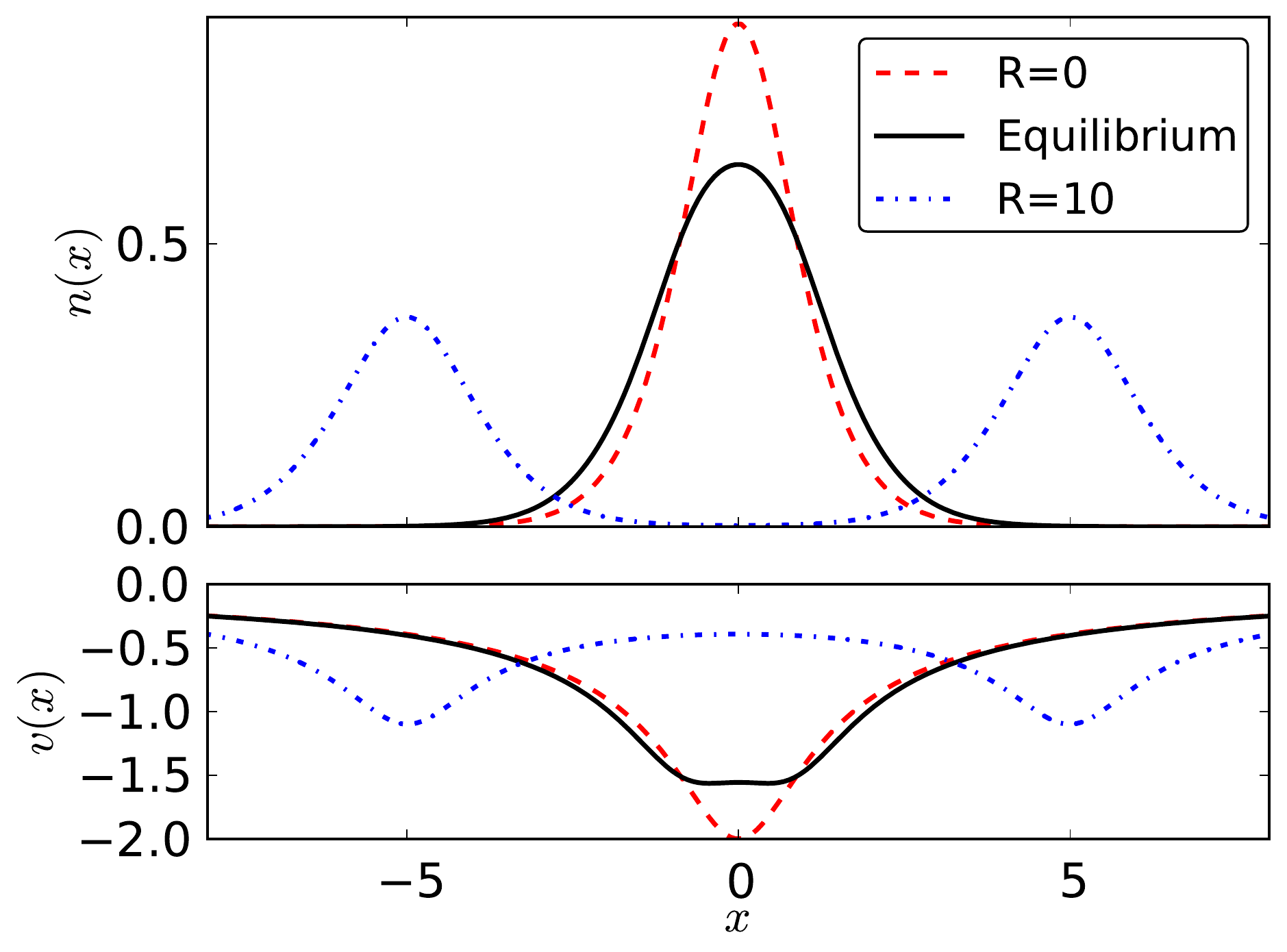}
\caption{The soft-Coulomb model of a diatomic for $Z=1$, $N=2$, which we test
our ML method on. 
External potentials and corresponding KS densities are shown
for $R=0$ (dashed), equilibrium bond length at R=1.62 (solid), 
and stretched at $R=10$ (dot-dashed), in atomic units.}
\label{fig:basic-1}
\end{figure}

Here, we consider a one-dimensional model of diatomic molecules, where
the electron repulsion has the soft-Coulombic form \cite{WSBW12}
\ben
v_{\rm ee}(u) = {1 \over \sqrt{1 + u^2}},
\een
as this has been used in a variety of contexts.
The one-body potential attraction of the two ``nuclei'' of nuclear charge $Z$ at separation $R$ is
\ben
v(x) = -Z(v\ee(x - R/2) + v\ee(x + R/2)),
\een
and the internuclear repulsion is $V_{\rm NN}(R) = Z^2v\ee(R)$.
We solve this model within KS-DFT \cite{KS65} with the
local density approximation (LDA) for XC \cite{HFCV11,WSBW12}.
The spin-unpolarized form of LDA exchange for this system is given in Ref. \cite{WSBW12},
and an accurate parametrization of the LDA correlation energy is given in Ref. \cite{HFCV11}.

Our goal is to ``learn'' the non-interacting kinetic energy $T\s[n]$ of the KS system.
In previous work \cite{SRHM12}, where we demonstrated for the first time the ability of ML
to approximate density functionals, the fermions were non-interacting and confined to live in a box, restraining
the variety of possible densities. In particular, there was no analog of 
a binding energy curve where a density is centered on two sites whose separation
varies continuously from small to infinite.

Fig.~\ref{fig:basic-1} shows the
densities and potentials for the united atom, equilibrium bond length, and stretched diatomic.
To generate a dissociation curve like that of Fig 1, we consider bond lengths
up to $R=10$, and place the entire system on 
a 500-point grid from $x=-20$ to 20. This is necessary to converge our KS-DFT calculations.
We doubly-occupy the lowest $Z$ orbitals, so that $N=2Z$, where $N$ is the
number of electrons.  We extract various energies and the density as a
function of $R$ for different values of $N$.

A curious point is that
the ML method only needs 50 points to achieve the level of accuracy given in this paper
{\em once the training energies are sufficiently accurate}.  Thus, a 500-point
calculation may be needed to find essentially exact energies, but
a far smaller grid (50 points) then yields a sufficiently accurate representation
of the densities for the ML to achieve arbitrarily accurate energies.
ML automatically corrects for the coarseness of the grid.

\renewcommand*\arraystretch{1.4}
\begin{table}[t]
\begin{tabular}{|c@{\hspace{0.8em}}c@{\hspace{0.8em}}c@{\hspace{0.6em}}l@{\hspace{1.6em}}l@{\hspace{1.6em}}l@{\hspace{1em}}l|}
\toprule[0.1em]
$N$ & $M$ & $\lambda$ & $\sigma$ & $\overline{| \Delta T |}$ & ${| \Delta T |}^{\text{std}}$ & ${| \Delta T |^{\text{max}} \over \overline{| \Delta T |}}$ \\
\toprule[0.1em]
2    & 4    & 1.4e-3       & 105.         & 20       & 15       & 2.2            \\
2    & 6    & 4.8e-3       & 5.2           & 2.9       & 3.6       & 4.1            \\
2    & 8    & 4.1e-5       & 39.3          & 2.6       & 2.8       & 3.7            \\
2    & 10   & 3.0e-7       & 32.5          & 0.13[0.93]       & 0.19[2.6]       & 6.1[14]            \\
2    & 12   & 5.6e-7       & 15.9          & 0.092       & 0.14       & 6.1            \\
2    & 15   & 2.4e-8       & 17.8          & 0.045       & 0.047       & 3.9            \\
2    & 20   & 2.4e-8       & 15.2          & 0.038       & 0.037       & 3.5            \\
2    & 25   & 1.9e-10       & 6.1           & 0.002       & 0.002       & 3.9            \\
\midrule[0.05em]
4    & 10   & 3.3e-7       & 87.7          & 0.25[1.3]       & 0.31[3.7]       & 4.9[15]            \\
4    & 20   & 1.1e-10       & 46.1          & 0.02       & 0.018       & 3.9            \\
\midrule[0.05em]
6    & 10   & 8.1e-6       & 74.5          & 3.1[13]       & 4.0[33]       & 4.7[11]            \\
6    & 20   & 1.8e-9       & 55.1          & 0.016       & 0.018       & 4.8            \\
\midrule[0.05em]
8    & 10   & 2.0e-6       & 56.2          & 1.4[7.7]       & 1.7[22]       & 5.0[14]            \\
8    & 20   & 2.9e-9       & 16.1          & 0.016       & 0.012       & 3.3            \\
%\midrule[0.05em]
%	1-4$^\dagger$ & 400 & $3.2$ & 47 & 0.12	& 0.20 		& 3.6 \\ 
\bottomrule[0.1em]
\end{tabular}
\caption{
Parameters and errors (in kcal/mol)
as a function of fermion number $N$ and number of training densities $M$.
Mean, standard deviation and max values taken over 199 test densities
evenly spaced from $R=0$ to 10 (inclusive). Brackets
represent errors on self-consistent densities.}
\label{tbl:modelerrors}
\end{table}

To construct the model, we choose $M$ training densities at evenly spaced $R$
between 0 and 10 (inclusive).
Table \ref{tbl:modelerrors} shows the performance of the MLA, evaluated on a test set
with $M=199$. To compare, we tested the
LDA in 1d, $T\loc[n] = \pi^2 \int dx\, n^3(x)/6$, and a modified gradient expansion approximation (GEA) \cite{LCPB09},
$T^{\rm MGEA}[n] = T\loc[n] + c\, T\W[n]$, where $c=0.559$ has been chosen to minimize the error.
LDA has a MAE of 45 kcal/mol, while the GEA only improves that to 33 kcal/mol. For $N=2$, 
we have already achieved a MAE below 1 kcal/mol for $M=10$.
No other existing approximations
achieve this level of accuracy, which can be systematically improved.

\begin{figure}[t]
\includegraphics[width=8cm]{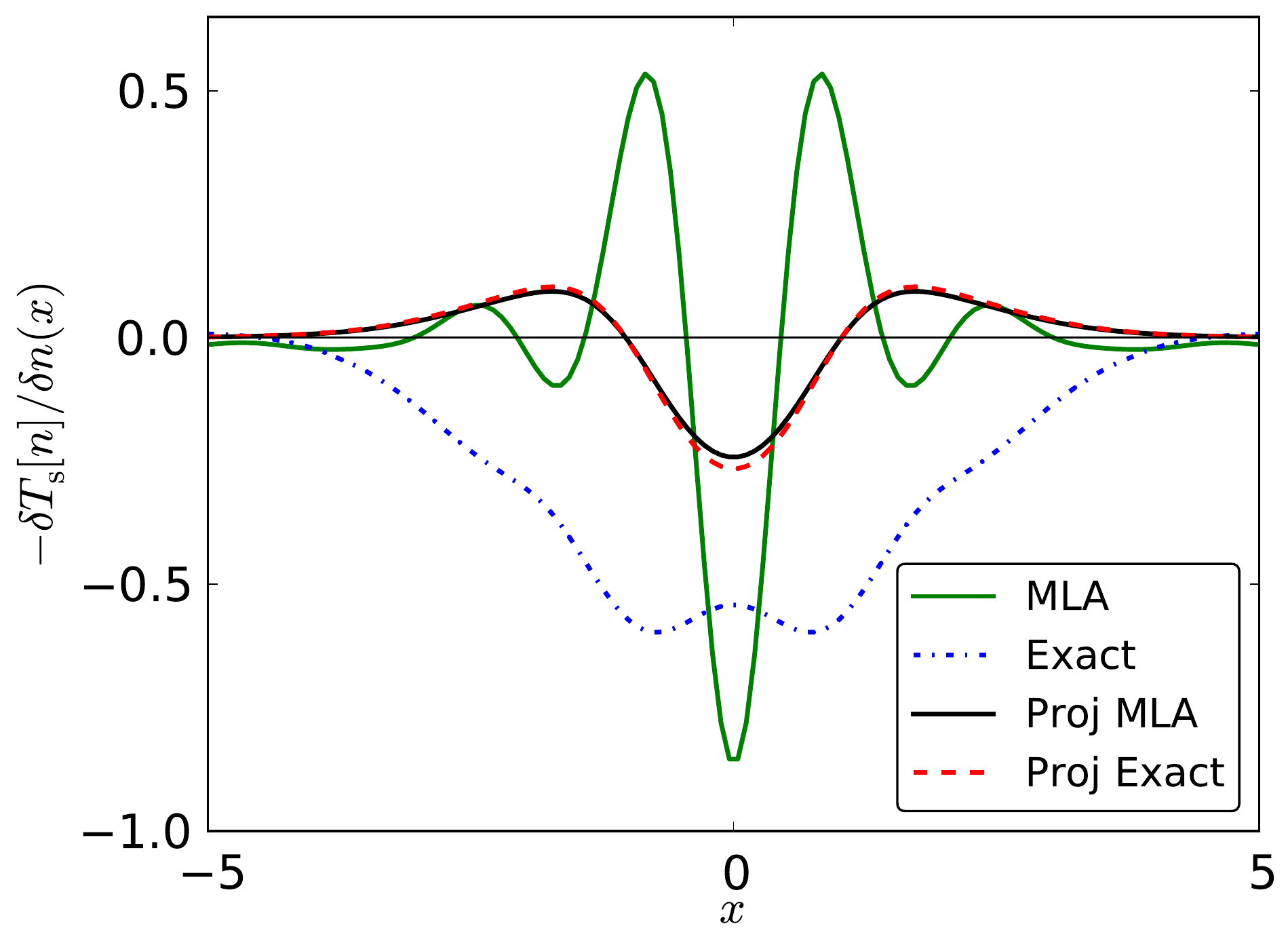}
\caption{The functional derivative of our MLA (green) cannot reproduce the exact
derivative (blue dot dashed, given by $-v\s[n]$ if evaluated at the ground-state density) because 
this information is not contained in the data. However, both agree when projected
onto the space of the data (black and red dashed). Shown for $Z=1$ and $M=10$ 
at equilibrium bond length (R=1.62), in atomic units.}
\label{fig:basic-2}	
\end{figure}

Thus far, our results have been reported evaluating the ML functional on exact
densities.  In real applications of orbital-free DFT, these are not available.   
The density that minimizes the approximate energy satisfies:
\ben
\frac{\delta T\s[n]}{\delta n(x)} = \mu- v\s(x),
\label{eq:totEmin}
\een
where $\mu$ is a constant.
But just as in Ref. \cite{SRHM12}, ML does {\em not}
yield an accurate functional derivative.
Fig.~\ref{fig:basic-2} shows that the derivative
of our MLA evaluated at the ground-state density is very different from the exact one.
The exact functional derivative shows how the KE changes along {\em any} direction, but the 
model only knows about directions along 
which the training densities lie. Since these densities are generated from a set of potentials
parametrized by $R$, the data is effectively 1d (locally), embedded in a high-dimensional space.
The same effect occurs in \cite{SRHM12}, except
there we had 9 parameters in the potential.

\begin{figure}[b]
\includegraphics[width=8.5cm]{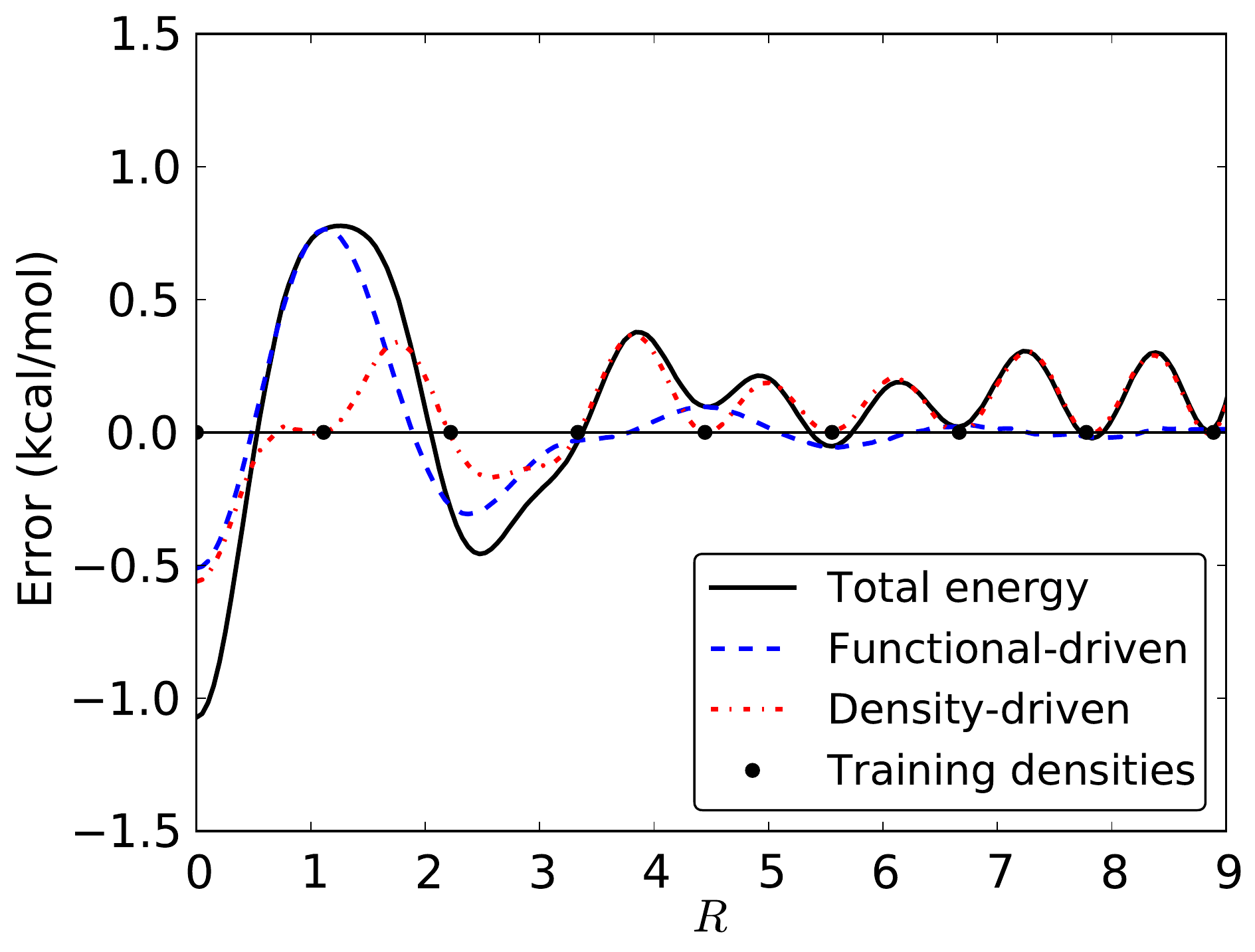}
\caption{The total error of the model and the functional- and density-driven errors \cite{KSB13}.
The dots mark the location of the training densities. $R$ given in atomic units, for $Z=1$ and 
$M=10$. 
}
\label{fig:calc-1}
\end{figure}

The bare gradient of the MLA makes solving Eq.~\ref{eq:totEmin} unstable.
If we project the gradient onto the space of the data, we can extract
the projected functional derivative of the MLA (see Fig.~\ref{fig:basic-2}).
In \cite{SRHM12}, we used principal component analysis (PCA) to 
locally approximate the neighborhood of training densities as linear, but 
here the local neighborhood is nonlinear, and the PCA method leads to inaccurate 
self-consistent densities.
Instead, we use the nonlinear gradient de-noising (NLGD) projection method
developed in Ref. \cite{SRHB13}.

\begin{figure}[t]
\includegraphics[width=7.5cm]{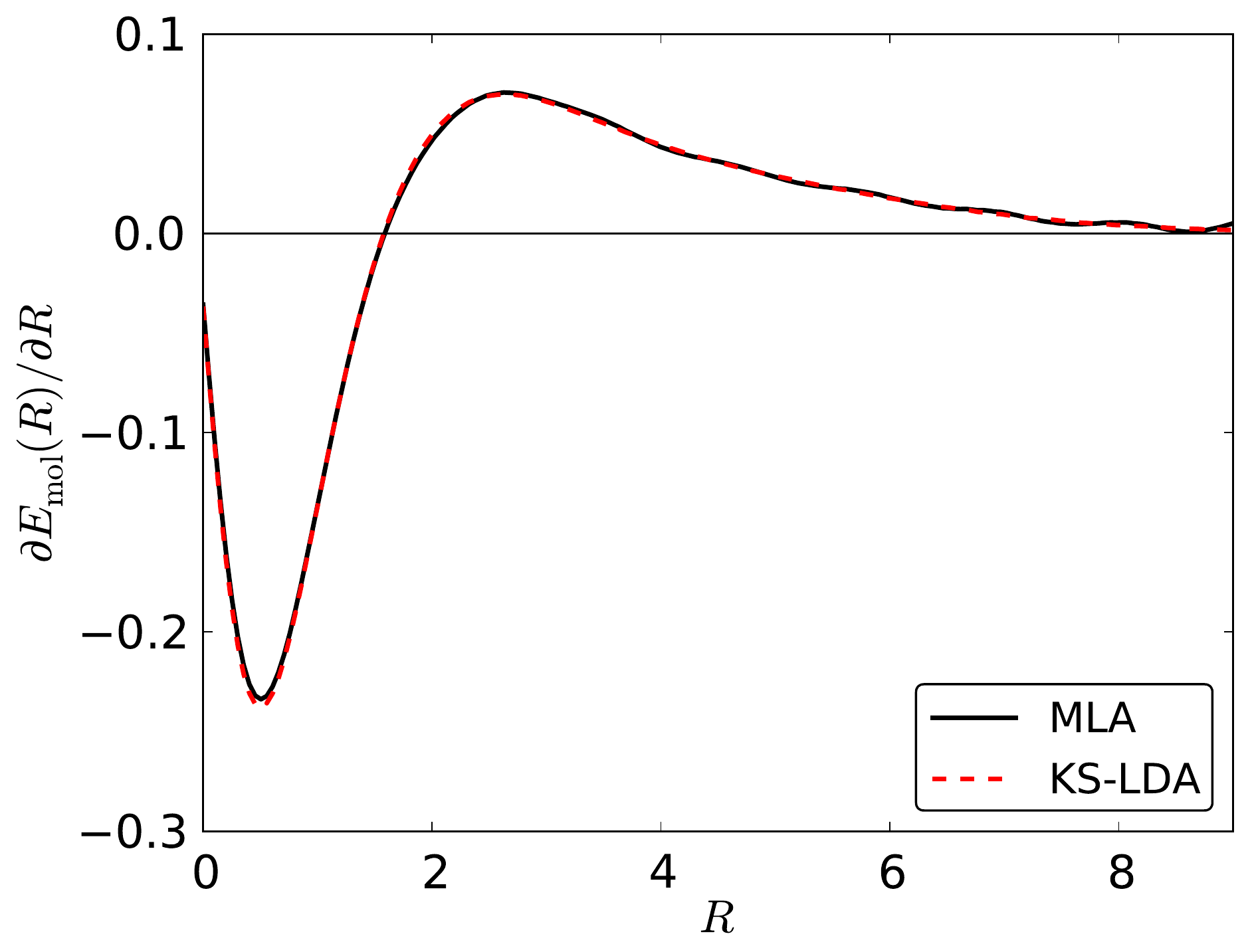}
\caption{Molecular forces as a function of $R$, for $Z=1$ and 10 training densities.
Derivatives are calculated via finite-difference.
}
\label{fig:scd-4}
\end{figure}

Applying the NLGD projected gradient descent technique \cite{SRHB13} to 
minimize Eq. \ref{eq:totEmin} with $\TML$
yields the results in Fig. \ref{fig:calc-1}, where we plot the error in total energy,
$\Delta E = E\ML[\tilde n] - E[n]$, where $n$ is the exact density and $\tilde n$
is the self-consistent density. This is split into the error due to the approximate
functional (i.e. functional-driven error),
$\Delta E_{\rm F} = E\ML[n] - E\LDA[n]$, and the error due to the deviation of
the self-consistent density from the exact (i.e. density-driven error),
$\Delta E_{\rm D} = E\ML[\tilde n] - E\ML[n]$ \cite{KSB13}. Near $R=0$,
$\Delta E_{\rm F}$ is larger because the $T\s$ is rapidly changing. $\Delta E_{\rm D}$
is zero at the training densities and is largest farthest from them. At large $R$,
the $\Delta E_{\rm D}$ dominates over $\Delta E_{\rm F}$.
Table~\ref{tbl:modelerrors} gives the MAE of the ML functional evaluated
on the self-consistent densities for different $N$ and $M$.

Fig.~\ref{fig:scd-4} shows the forces calculated with model densities.
The forces are very accurate and
should be suitable for, e.g., an {\it ab-initio} molecular dynamics calculation.
Combined with the reduced computational cost of orbital-free DFT, this method has the potential to
simulate very large systems at the same level of accuracy KS-DFT currently provides.

Our work shows that, for a one-dimensional system, machine
learning can be used to find a density functional that produces accurate
energies and forces on self-consistent densities, even when bonds break.
No existing orbital-free scheme comes close to this level of accuracy.
Although this example is limited to one dimension, there is no reason
in principle to doubt the efficacy of the method for real bonds.

The authors thank IPAM
at UCLA for hospitality and acknowledge NSF Grant No. CHE-1240252 (JS, KB),
EU PASCAL2 and DFG Grant No. MU 987/4-2 (K.H., and K. R. M.), EU Marie
Curie Grant No. IEF 273039 (M. R.), and NRF Korea Grant No. R31-10008 (K. R. M.).

%\bibliography{ML}

\end{document}